\documentclass[mathleft]{an}
\usepackage{graphicx}

\usepackage{times}
\begin{document}

\Pagespan{1}{}
\Yearpublication{2011}%
\Yearsubmission{2011}%
\Month{9}%
\Volume{332}%
\Issue{9/10}%
 \DOI{10.1002/asna.201111599}%

\title{Spectroscopic variability and magnetic fields of HgMn stars%
\thanks{Based on ESO Archival data, from ESO programme 083.D-1000(A).}}

\author{S.~Hubrig\inst{1}\fnmsep\thanks{Corresponding author:
  {shubrig@aip.de}}
  \and J.F.~Gonz\'alez\inst{2}
  \and I.~Ilyin\inst{1}
  \and H.~Korhonen\inst{3,~4}
  \and I.S.~Savanov\inst{5}
  \and T.~Dall\inst{6}
  \and M.~Sch{\"o}ller\inst{6}
  \and C.R.~Cowley\inst{7}
  \and M.~Briquet\inst{8}\fnmsep\thanks{Postdoctoral Fellow of the Fund for Scientific Research of Flanders (FWO), Belgium.}
  \and R.~Arlt\inst{1}
}

\titlerunning{Variability and magnetic fields of HgMn stars}
\authorrunning{S. Hubrig et al.}
\institute{
Leibniz-Institut f{\"u}r Astrophysik Potsdam (AIP), An der Sternwarte~16, 
14482 Potsdam, Germany
\and 
Instituto de Ciencias Astronomicas, de la Tierra, y del Espacio (ICATE), 5400 
San Juan, Argentina
\and
Niels Bohr Institute, University of Copenhagen, Juliane Maries Vej~30, 2100 
Copenhagen, Denmark 
\and
Finnish Centre for Astronomy with ESO (FINCA), University of Turku, V{\"a}is{\"a}l{\"a}ntie 20, FI-21500 Piikki{\"o}, Finland
\and
Institute of Astronomy, Russian Academy of Sciences, Pyatnitskaya 48, Moscow 119017, Russia 
\and
European Southern Observatory, Karl-Schwarzschild-Str.~2, 85748 Garching bei M\"unchen, Germany       
\and
Department of Astronomy, University of Michigan, Ann Arbor, MI, 48109-1042, USA
\and 
Instituut voor Sterrenkunde, Katholieke Universiteit Leuven, Celestijnenlaan 
200~D, 3001 Leuven, Belgium
}

\received{2011 Sep 23}
\accepted{2011 Sep 29}
\publonline{2011}


\abstract{The discovery of exotic abundances, chemical inhomogeneities, and weak 
magnetic fields on the surface of late B-type primaries in spectroscopic 
binaries has important implications not only for our understanding of 
the formation mechanisms of stars with Hg and Mn peculiarities themselves, 
but also for the general understanding of B-type star formation in binary systems. 
The origin of the abundance anomalies observed in late B-type stars with HgMn peculiarity is still poorly 
understood. The connection between HgMn peculiarity and membership in binary and multiple systems is supported by our 
observations during the last decade. The important result achieved in our studies of a large sample of HgMn stars is the 
finding that most HgMn stars exhibit spectral variability of various chemical elements, proving that the presence of an 
inhomogeneous distribution on the surface of these stars is a rather common characteristic and not a rare phenomenon. 
Further, in the studied systems, we found that all components are chemically peculiar with different abundance patterns. 
Generally, He and Si variable Bp stars possess large-scale organised magnetic fields that in many cases appear to occur 
essentially in the form of a single large dipole located close to the centre of the star. The presence of magnetic fields 
in the atmospheres of HgMn stars has been demonstrated in several studies. In addition to the measurements of 
longitudinal and quadratic magnetic 
fields, this work also showed evidence for a relative magnetic intensification of Fe\,{\sc ii} lines produced by different 
magnetic desaturations induced by different Zeeman-split components.
}
\maketitle

\section{Introduction}

Observational data suggest that approximately two-thirds of all solar-type 
field stars form in binary, triple, or higher-order systems and that all 
massive stars are in multiple systems, preferentially in higher-order 
systems rather than in binaries (e.g.,  Duquennoy \& Mayor \cite{Duquennoy1991}; 
Tokovinin \& Smekhov \cite{Tokovinin2002}; Kobulnicky \& Fryer \cite{Kobulnicky2007}). 
Studies of stars of various mass in binary systems are of particular interest for a number of reasons.
Binarity makes it possible to obtain solid data on stellar fundamental properties 
such as mass and radius as well as the evolutionary status of both components. Furthermore, the analysis of
the chemical composition of the components gives clues to the origin of their frequently observed  
chemical surface anomalies. 

Over the last years, we have performed 
extensive spectroscopic studies of upper-main sequence spectroscopic binaries with 
late B-type primaries  (spectral types B7--B9)
with the goal to understand why the vast majority of these stars exhibits  
certain chemical abundance anomalies, i.e. large excesses of P, Mn, Ga, Br, Sr, Y, Zr, 
Rh, Pd, Xe, Pr, Yb, W, Re, Os, Pt, Au, and Hg, and 
underabundances of He, Al, Zn, Ni, and Co (e.g. Castelli \& Hubrig \cite{Castelli2004a}). Strong isotopic 
anomalies were detected for the chemical elements Ca, Pt, and Hg with patterns changing from one star to 
the next (Hubrig et al.\ \cite{Hubrig1999a}; Dolk et al.\ \cite{Dolk2003}; 
Castelli \& Hubrig \cite{Castelli2004b}; Cowley et al.\ \cite{Cowley2008}).
The presence of weak emission lines of various elements was for the first time reported by 
Wahlgren \& Hubrig (\cite{Wahlgren2000}).
Observationally, these stars are characterized by low rotational 
velocities ($\langle v\sin i\rangle \leq 29$~km\,s$^{-1}$, 
Abt et al.\ \cite{Abt1972}). The fraction of these chemically peculiar stars, 
usually called HgMn stars, decreases with increasing rotational velocity. Evidence that stellar rotation does 
affect abundance anomalies in HgMn stars is provided by the rather 
sharp cutoff in such anomalies at a projected rotational velocity 
of 70--80~km\,s$^{-1}$ (Hubrig \& Mathys \cite{Hubrig1996}).

More than 2/3 of the HgMn stars are known to belong to spectroscopic 
binaries (Hubrig \& Mathys \cite{Hubrig1995}) with a preference of orbital 
periods in the range between 3 and 20 days.
It is striking that the inspection of SB systems with a late B-type 
primary in the 9th Catalogue of Spectroscopic Binary 
Orbits (Pourbaix et al.\ \cite{Pourbaix2004}) indicates a strong correlation 
between the HgMn peculiarity and membership in a binary system:
among bright well studied SB systems with late B-type slowly rotating 
\mbox{($v\sin i<\,70~{\rm km\,s}^{-1}$)} primaries with an apparent 
magnitude of up to $V\approx7$  and orbital periods between 3 and 
20\,days, apart from HR\,7241, all 21~systems have a primary with 
a HgMn peculiarity. Based on this fact, it is very likely that the 
majority of slowly rotating late B-type stars formed in binary 
systems with certain orbital parameters become HgMn stars, indicating that careful studies of these peculiar 
stars are important for the general understanding of B-type star 
formation in binary systems. Since a number of HgMn stars in
binary systems is found at the zero-age main sequence (ZAMS)
(e.g., Nordstrom \& Johansen \cite{Nordstrom1994}; Gonz\'alez et al.\ \cite{Gonzalez2010}), it is expected
that the timescale for developing a HgMn peculiarity is very 
short.
  
A large number of HgMn stars belong to triple
or even quadruple systems
according to speckle interferometry, diffraction-limited near-infrared imaging
with NAOS-CONICA at the VLT (Cole et~al.\ \cite{Cole1992}; Isobe \cite{Isobe1991}; Sch\"oller et al.\ \cite{Scholler2010}) 
and observations of X-ray emission that appears to always come from
a cool companion (Hubrig \& Bergh\"ofer \cite{Hubrigetal1998}; Hubrig et al. \cite{Hubrig2001a}).

HgMn stars were assumed in the past not to possess magnetic fields or to exhibit
spectral line variability such as commonly shown by chemically peculiar magnetic Ap and Bp stars.
As more than 2/3 of the HgMn stars are known to belong to spectroscopic 
binaries, the variation of spectral lines observed in any HgMn star is usually 
explained to be due to the orbital motion of the companion. 
The aspect of inhomogeneous distribution of some 
chemical elements over the surface of HgMn stars was, for the first time, discussed by Hubrig \& Mathys 
(\cite{Hubrig1995}). 
From a survey of HgMn stars in close SBs, it was suggested that some chemical elements might be 
inhomogeneously distributed on the surface, with, in particular, preferential concentration of Hg along 
the equator. In close SB2 systems where the orbital plane has a 
small inclination to the line of sight, 
a rather large overabundance of Hg was found. By contrast, in stars with orbits almost perpendicular to 
the line of sight, mercury is not observed at all. 

The first definitively identified spectrum variability 
which is not caused by the companion was reported for the binary HgMn star $\alpha$\,And by Wahlgren, 
Ilyin \& Kochukhov (\cite{Wahlgren2001}) and Adelman et al.\ (\cite{Adelman2002}). They suggested that the 
spectral variations of the 
Hg\,{\sc ii} line at $\lambda$3984 \AA\ discovered in high-dispersion spectra are not due to the orbital motion of the 
companion, but produced by the combination of the \hbox{2.8-d} period of rotation of the primary and a non-uniform 
surface distribution of mercury which is concentrated in the equatorial region, in good correspondence 
with the results of Hubrig \& Mathys (\cite{Hubrig1995}).  The variability of the Hg\,{\sc ii} line at $\lambda$3984 \AA\ was 
interpreted with a Doppler Imaging code revealing high-contrast mercury spots located along the 
rotational equator. Using Doppler Imaging reconstruction of spectroscopic time series obtained 
over seven consecutive years, Kochukhov et al.\ (\cite{Kochukhov2007}) suggested the presence of a secular 
evolution of the mercury distribution.

Importantly, recent results (e.g., Nu{\~n}ez, Gonz\'alez \& Hubrig \cite{Nunez2011}) show that line profile variability is a 
general characteristic 
of HgMn stars, rather than an exception. This variability is caused by an inhomogeneous chemical element distribution, and 
implies that most HgMn stars present a non-uniform distribution of one or more chemical elements. 

\section{Spectroscopic variability as a general characteristics of HgMn stars}

\begin{figure*}
\centering
\includegraphics[width=0.99\textwidth]{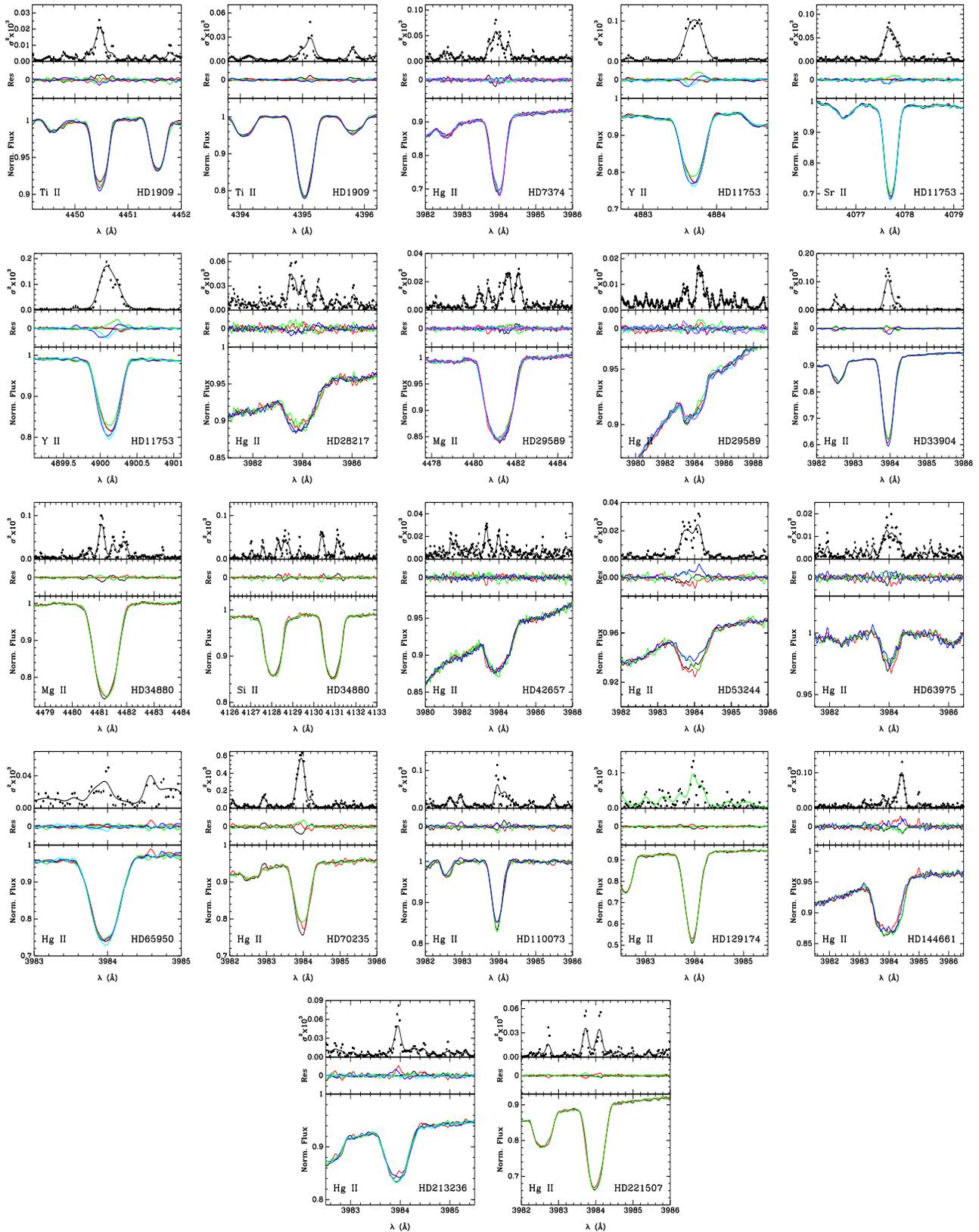}
\caption{(online colour at: www.an-journal.org) Examples of line profile variability in various HgMn stars. In each plot the middle panel has the 
same scale as the lower panel. The curves in the upper panels present a smoothing
of the residual points using convolution with a Gaussian of $\sigma$ = 2.05 pixels.}
\label{variability}
\end{figure*}

As a result of our extensive spectroscopic studies in the last years,
we could establish that chemical inhomogeneities are widespread among 
B7--B9 HgMn primaries of spectroscopic binaries and that previous failures 
to detect them were largely related to a small number of repeated 
observations of the same targets and traditional focus on the sharp-lined 
stars, for which the spectrum variations are much harder to detect. 
High-quality spectra of a representative sample of 
HgMn stars were obtained with UVES (Ultraviolet and Visual Echelle Spectrograph) at the VLT 
and the Fiberfed Extended Range Optical Spectrograph (FEROS) 
at the ESO 2.2-m telescope within the framework of our ESO 
programs aimed at a careful study of line profile variations 
of various elements. 
In Fig.~\ref{variability} we present numerous examples of the variability of various spectral lines
belonging to different elements in HgMn stars.

\begin{figure*}
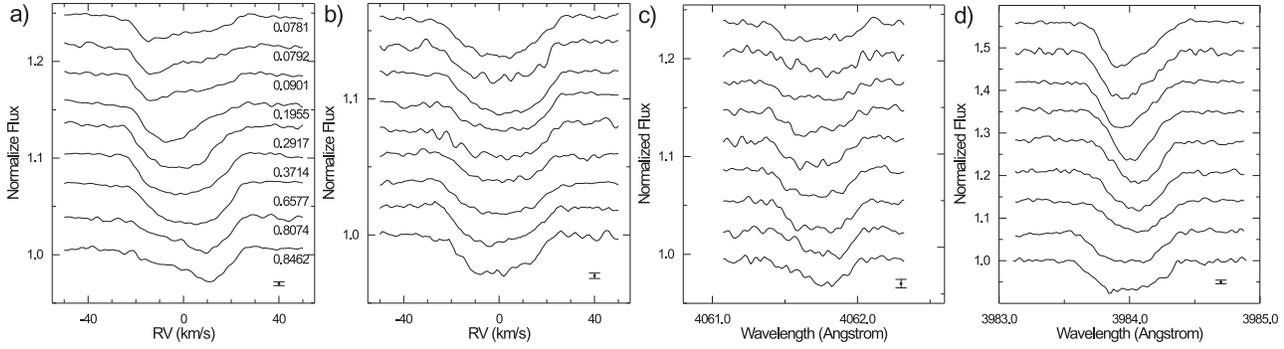

\centering
\includegraphics[width=0.24\textwidth]{fig1a-Yphase.eps}
\includegraphics[width=0.24\textwidth]{fig1b-Zr.eps}
\includegraphics[width=0.24\textwidth]{fig1c-Pt.1.eps}
\includegraphics[width=0.253\textwidth]{fig1d-Hg.eps}
\caption{Variations of line profiles in spectra of the zero-age main-sequence eclipsing 
binary AR\,Aur phased with the rotation period $P=4.13$\,days:
\emph{a})~Y\,{\sc ii},
\emph{b})~Zr\,{\sc ii},
\emph{c})~Pt\,{\sc ii} $\lambda\,4061.7$ \AA,
\emph{d})~Hg\,{\sc ii} $\lambda\,3983.9$ \AA.
The rotational phase increases from \emph{top to bottom} -- see \emph{a}). Error bars in the 
lower corner on the right side indicate the standard error of 
the line profiles. The spectra are shifted in vertical direction for display purposes.}
\label{fig1}  
\end{figure*}

Using nine high signal-to-noise high-resolution UVES spectroscopic observations of the eclipsing SB2 HgMn binary star AR\,Aur
it was soon demonstrated that the spots of $\alpha$\,And are not unique
(Hubrig et al.\ \cite{Hubrig2006a}).
The zero-age main-sequence (ZAMS) eclipsing binary AR Aur (HD\,34364, B9V+B9.5V) with an orbital 
period of 4.13\,d at an age of only  $4{\times}10^6$\,years belongs to the Aur OB1 association 
(Nordstrom \& Johansen \cite{Nordstrom1994}) 
and presents the best case for a study of evolutionary aspects of the chemical peculiarity phenomenon.
The problem of 
analysing the component spectra in double-lined spectroscopic binaries 
is difficult, but, fortunately, in the past few years several techniques 
for spectral disentangling have been developed. For each observed phase 
we applied the procedure of decomposition described in detail by 
Gonz\'alez \& Levato (\cite{Gonzalez2006}).
In Fig.~\ref{fig1}, we show the behavior of the line profiles of a few
elements over the rotation period. We found that Zr\,{\sc ii}, Nd\,{\sc iii}, Pt\,{\sc ii}, 
and He\,{\sc i} lines appear rather weak, but still their variations are definite.
Interestingly, while the behavior of the line profiles of 
Y\,{\sc ii}, Pt\,{\sc ii}, Hg\,{\sc ii}, Sr\,{\sc ii}, and Nd\,{\sc iii} is rather similar over the 
rotation period, the line profiles of Zr\,{\sc ii} and He\,{\sc i} seem to vary with 
a 180$^\circ$ phase shift. 

\begin{figure*}
\centering
\includegraphics[width=0.81\textwidth]{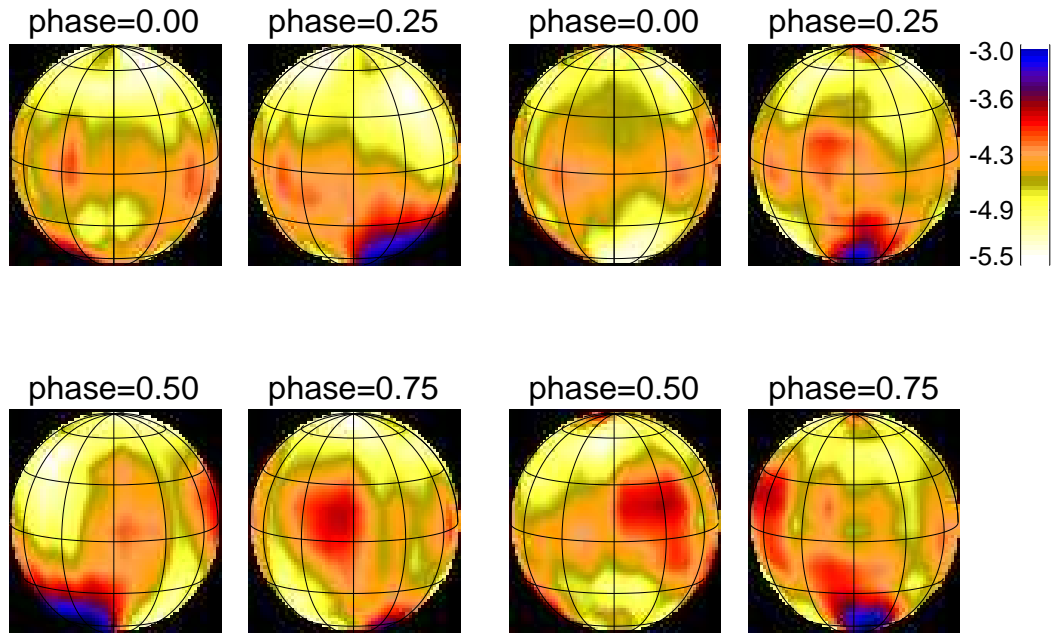}
\caption{(online colour at: www.an-journal.org) The Fe abundance map of AR\,Aur obtained from the Fe\,{\sc ii} 4923.9\,\AA{}
line for SET1 (\emph{left}) and SET2 (\emph{right}). }
 \label{fig:Femap}
\end{figure*}

\begin{figure*}
\centering
\includegraphics[width=0.81\textwidth]{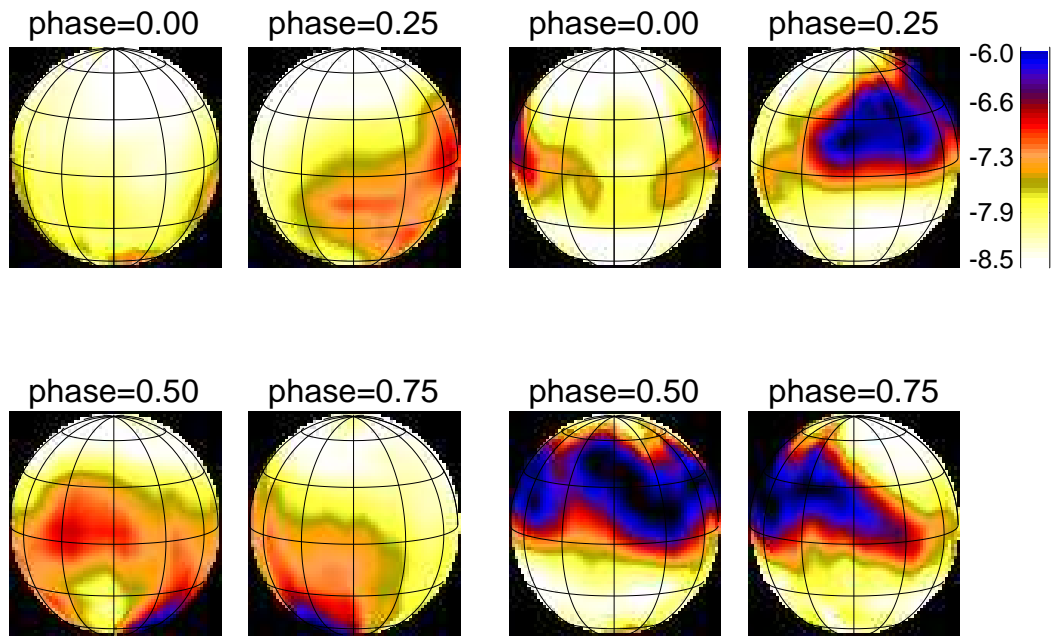}
\caption{(online colour at: www.an-journal.org) The Y abundance map of AR\,Aur obtained from the Y\,{\sc ii} 4900.1\,\AA{}
line for SET1 (\emph{left}) and SET2 (\emph{right}).}
 \label{fig:Yemap}
\end{figure*}

First Doppler maps for the elements Mn, Sr, Y, and Hg using 
nine spectra of AR\,Aur observed with the UVES spectrograph in 
2005 were for the first time presented at the IAU Symposium~259 
by Savanov et al.\ (\cite{Savanov2009}).
In the most recent analysis of this system (Hubrig et al.\ \cite{Hubrig2010}) based 
on the spectra obtained with the Coud\'e 
Spectrograph of the 2.0-m telescope of the Th\"uringer Landessternwarte (TLS;
Oct.\ 2008--Feb.\ 2009) and the SES spectrograph 
of the \hbox{1.2-m} STELLA-I robotic telescope at the Teide Observatory (Nov.--Dec.\ 2008), 
we used the improved Doppler imaging 
code IA introduced by Freyhammer et al.\ (\cite{Freyhammer2009}). This code uses
Tikhonov regularization in a way similar to the Doppler Imaging 
(DI) method described by Piskunov (\cite{Piskunov2008}) with a grid of 6$^{\circ}\times$6$^{\circ}$. 
In the DI reconstruction, we search for the minimum of the regularized
discrepancy function, which includes the regularization 
function and the discrepancy function describing the difference 
between observed and calculated line profiles. For the Doppler 
Imaging reconstruction, we selected the elements Fe and Y with 
the clean unblended spectral lines Fe\,{\sc ii} 4923.9\,\AA{} and Y\,{\sc ii} 4900.1\,\AA{}, and 
showing distinct variability over the rotation period. 
The results of the reconstruction using UVES spectra (SET1) and most recent spectra obtained with smaller
telescopes (SET2) are presented in Figs.~\ref{fig:Femap} and \ref{fig:Yemap},
respectively.

The inspection of the resulting Fe and Y distribution maps 
separated by four years shows that Fe is overabundant by up to 
+1.5\,dex and Y is overabundant by up to +3.9\,dex in several spots. 
The positions and the shape of the spots with the highest Fe
overabundance slightly changed from 2005 to 2009, and the level of 
the Fe overabundance shows a significant increase, especially in 
the spot located close to the equator at the phases 0.50--0.75 and
in the polar spot at the phases 0.75--0.83. In the Y maps, the evolution
of overabundance, shape, and position of the spots appears 
much more remarkable, revealing a region of huge overabundance 
having a shape of a belt, which is broken around phase 0.
Intriguingly, in this phase we observe the hemisphere which is
permanently facing the secondary. Such a behavior is likely observed 
also for Sr in the UVES spectra and was discussed in our previous 
study (Hubrig et al.\ \cite{Hubrig2006a}).

For one SB1 system with a well pronounced variability, the HgMn star HD\,11753, it was possible to gather a 
large number of spectra with the CORALIE \'echelle spectrograph attached 
to the 1.2-m Leonard Euler telescope on La Silla in Chile (Briquet et al.\ \cite{Briquet2010}). In total, 
we obtained 113~spectra at a spectral resolution of 50\,000.
The radial velocities and equivalent widths were found to vary with the 
period $P=9.54$\,d.
We used Doppler imaging technique to 
reconstruct surface distribution of Ti, Sr, and Y. Ti and Y have numerous 
transitions in the observed optical spectral region allowing us to select 
unblended spectral lines with not too different line formation depths. 
The relevance of vertical abundance 
stratification in HgMn stars was previously discussed by  
Savanov \& Hubrig (\cite{Savanov2003}). 

\begin{figure}
\includegraphics[width=0.487\textwidth]{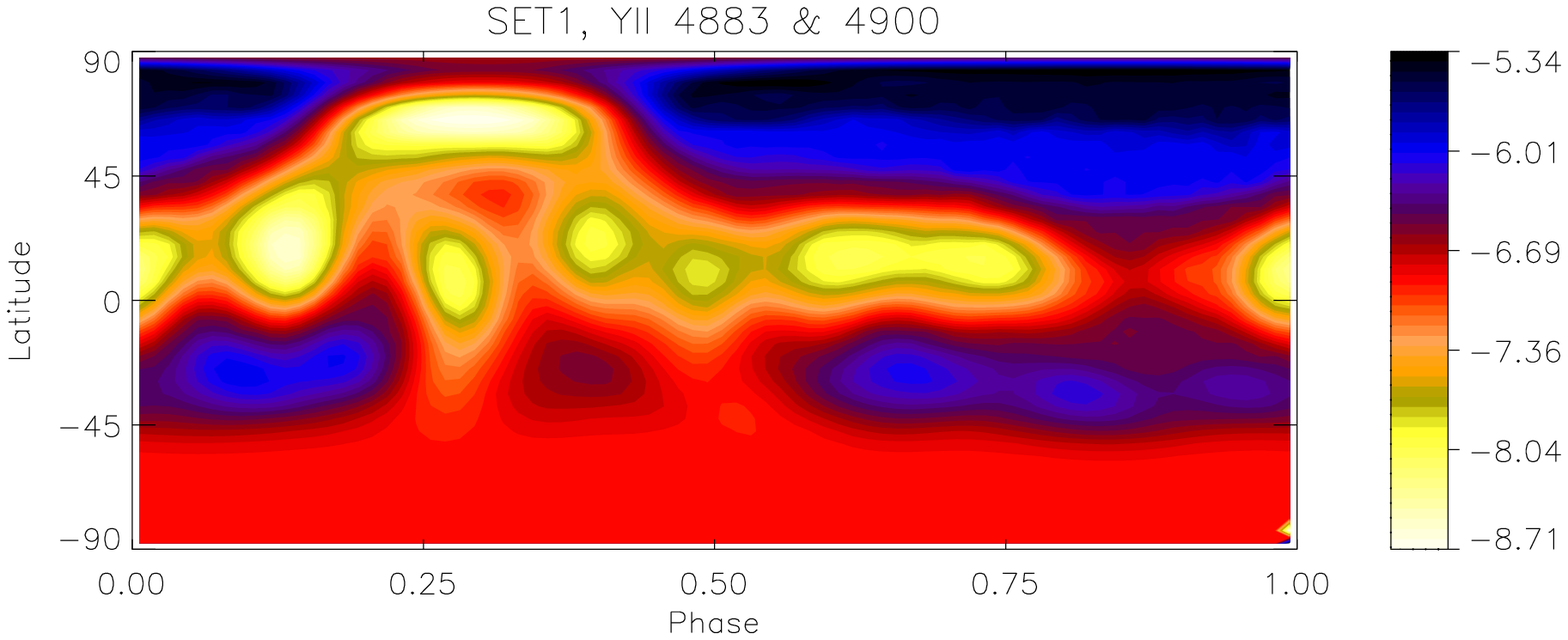}
\includegraphics[width=0.487\textwidth]{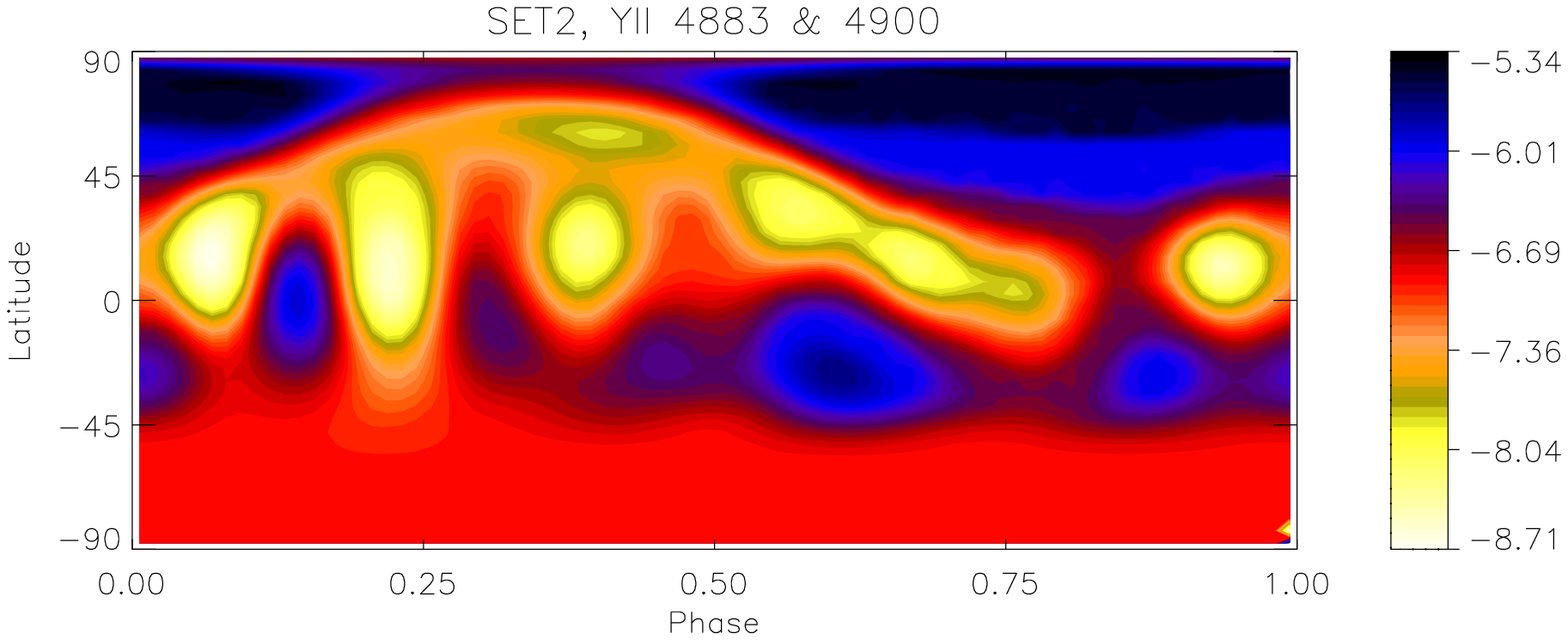}
\caption{(online colour at: www.an-journal.org) The Y abundance map of HD\,11753 obtained for SET1 (\emph{upper panel}) and SET2 (\emph{lower panel}). 
  The color indicates the abundance with respect to the total 
  number density of atoms and ions. }
 \label{fig:YIImap}
\end{figure}

The two sets of observations of HD\,11753 obtained in 2000 September 28\,--\,October 11 
(SET1) and in 2000 December 02\,--\,15 (SET2) consist of 76 and 28 observations, 
respectively, evenly spread over the stellar rotation cycle.
Surprisingly, results of Doppler imaging reconstruction revealed noticeable changes in the surface 
distributions of Ti\,{\sc ii}, Sr\,{\sc ii}, and Y\,{\sc ii} between the datasets separated by just 65 days,
indicating the presence of the hitherto not well understood physical processes in stars with radiative envelopes 
causing rather fast dynamical chemical spot evolution, not of the order of years, but rather months. 
We assume here that the features are real, and not an artifact of the
methodology.
All Ti, Sr, and Y abundance maps reveal a structure reminiscent of broken rings of low and high
abundance. 
As an example, the maps obtained from Y\,{\sc ii} lines 4883\,\AA{} and 4900\,\AA{} 
are presented
in Fig.~\ref{fig:YIImap}. The maps show broken abundance rings with the high 
abundance region extending from the latitude $45^{\circ}$ to the pole. The Y 
abundance distribution shows high latitude lower abundance spot around phases 
0.2--0.4, similarly to the Ti abundance maps. The equatorial region is dominated 
by a belt of lower abundance spots, and the regions below the equator by high 
abundance features. The average abundance of the Y maps is $-$6.77, which is 
significantly higher than the solar abundance of $-$9.80. We note that all the 
features revealed in the maps show abundances that are higher than the solar 
abundance of Y. Similar to the Ti maps, we observe in the Y maps that the lower
abundance high latitude feature at phases 0.2--0.4 becomes much less prominent 
in SET2.

The abundance maps of HD\,11753 presented in the work by Briquet et al.\ (\cite{Briquet2010}) exhibit 
clear differences between the surface abundance distribution of Ti, Sr, and Y. 
We also detect distinct differences in the spot configurations obtained from 
the same lines for different data sets, indicating a rather fast dynamical evolution of
the abundance distribution with time. 
Different dynamical processes take place in stellar radiation zones. An interaction 
between the differential rotation, 
the magnetic field, and the meridional circulation could possibly play a role in the generation of 
dynamical evolution of 
chemical spots. From the comparison of maps we find that it is possible that the Y distribution 
shows indications of increasing rotation rate towards the rotation pole, so-called differential 
rotation of anti-solar type.
On the other hand, these results are not sufficient to claim the presence of a surface 
differential rotation and further analyses of the elemental surface distribution in a larger sample 
of HgMn stars should be carried out before the implication of these new results can be discussed in more detail.

\section{Magnetic fields of HgMn stars}

Based on spectropolarimetric studies carried out during more than a dozen of years,
it now became clear that weak magnetic fields do exist in a number of HgMn stars
(Hubrig et al. \cite{Hubrig2008}, and references therein).
The most widespread method to detect a magnetic field is to obtain 
polarimetric spectra recorded in left- and right-hand polarized 
light to measure the mean longitudinal magnetic field. Another 
approach to study the presence of magnetic fields in upper main 
sequence stars is to determine the value of the mean quadratic 
magnetic field,
\begin{displaymath}
\langle B_{\rm q}\rangle= (\langle B^2\rangle + \langle B_{\rm z}^2\rangle)^{1/2},
\end{displaymath}
which is derived through the application of the moment technique, 
described by, e.g., Mathys \& Hubrig (\cite{Mathys2006}). Here $\langle B^2\rangle$ is 
the mean square magnetic field modulus (the average over the stellar 
disc of the square of the modulus of the field vector, weighted by 
the local emergent line intensity), while $\langle B_{\rm z}^2\rangle$ is 
the mean square longitudinal field (the average over the stellar 
disc of the square of the line-of-sight component of the magnetic 
vector, weighted by the local emergent line intensity). 

The mean quadratic magnetic field is determined from the study of the 
second-order moments of the line profiles recorded in unpolarized 
light (that is, in the Stokes parameter $I$). 
The second-order moment $R_I^{(2)}(\lambda_I) $ of a spectral 
line profile recorded in unpolarized light about its center of 
gravity $\lambda_I$ is defined as
\begin{displaymath}
R_I^{(2)}(\lambda_I) ={1\over W_\lambda}~\int{r_{{\cal
F}_I}(\lambda)(\lambda-\lambda_I)^2}~{\rm d}\lambda .
\end{displaymath}
The integration runs over the whole width of the observed 
line (see Mathys \cite{Mathys1988} for details). ${W_\lambda}$ is the 
line equivalent width; $r_{{\cal F}_I}$ is the line profile:
\begin{displaymath}
r_{{\cal F}_I}
=1-({\cal F}_I/{\cal F}_{I_{\rm c}}) .
\end{displaymath}

 $\ensuremath{{\cal F}_I} $ (resp. $\ensuremath{{\cal F}_{I_{\rm c}}} $)
is the integral over the visible stellar disk of the emergent intensity
in the line (resp. in the neighboring continuum).
The analysis is usually based on consideration of samples 
of reasonably unblended lines of Fe\,{\sc i} and Fe\,{\sc ii}. Using this method, 
Mathys \& Hubrig (\cite{Mathys1995}) could demonstrate the presence of quadratic 
magnetic fields in two close double-lined systems with HgMn primary 
stars, 74 Aqr and $\chi$ Lup.

Another method which can be applied to study magnetic fields in HgMn 
stars employs the relative magnetic intensification of the two Fe\,{\sc ii}
lines of multiplet 74, $\lambda\,6147.7$ and $\lambda\,6149.2$ \AA,
which have the same equivalent width to within 2.5\,\% in non-magnetic 
late B and A-type stars, but very different Zeeman patterns 
(Lanz \& Mathys \cite{Lanz1993}; Hubrig et al.\ \cite{Hubrig1999b};  Hubrig \& Castelli \cite{Hubrig2001b}). 
The observed discrepancy between the equivalent widths of these two 
lines in magnetic stars is attributed to magnetic intensification.
As the relative intensification is roughly correlated with the strength 
of the magnetic field, it is a powerful tool for detecting magnetic 
fields which have a complex structure and are difficult to analyze by 
polarization measurements. For a few HgMn stars, Hubrig \& Castelli (\cite{Hubrig2001b}) 
showed evidence for a relative magnetic intensification of Fe\,{\sc ii} lines 
of multiplet 74,
produced by different magnetic desaturations induced by different 
Zeeman-split components. 

\begin{figure*}
\vskip-3mm
\hskip-3mm
\includegraphics[width=0.50\textwidth]{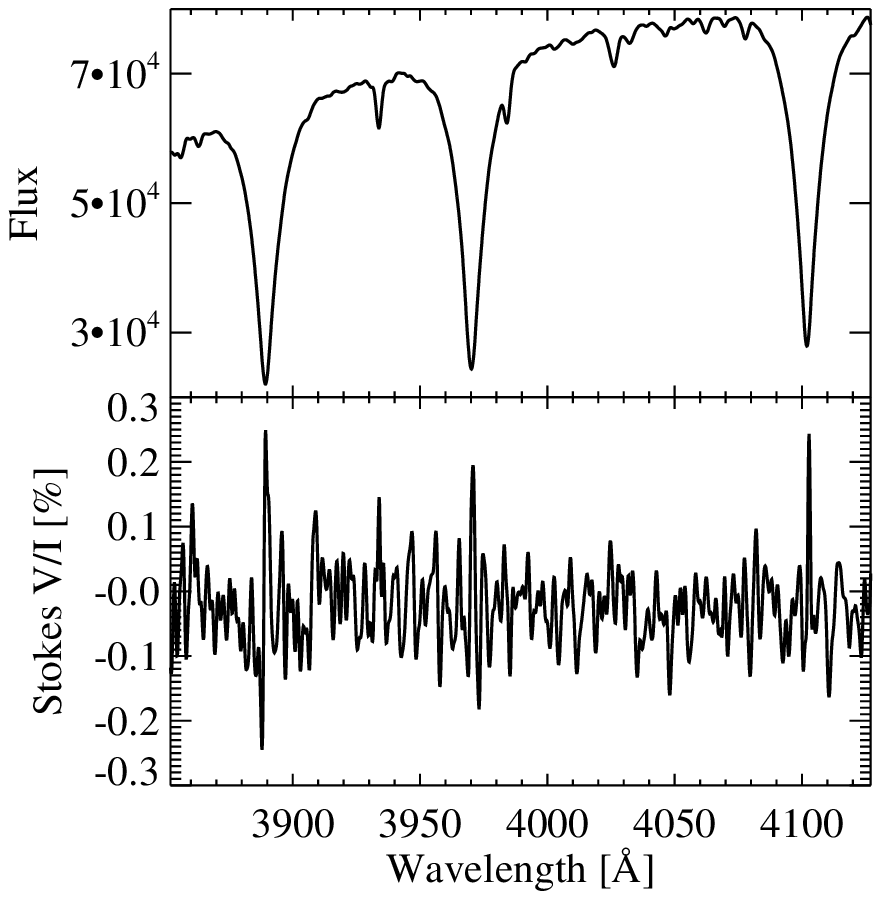}
\includegraphics[width=0.50\textwidth]{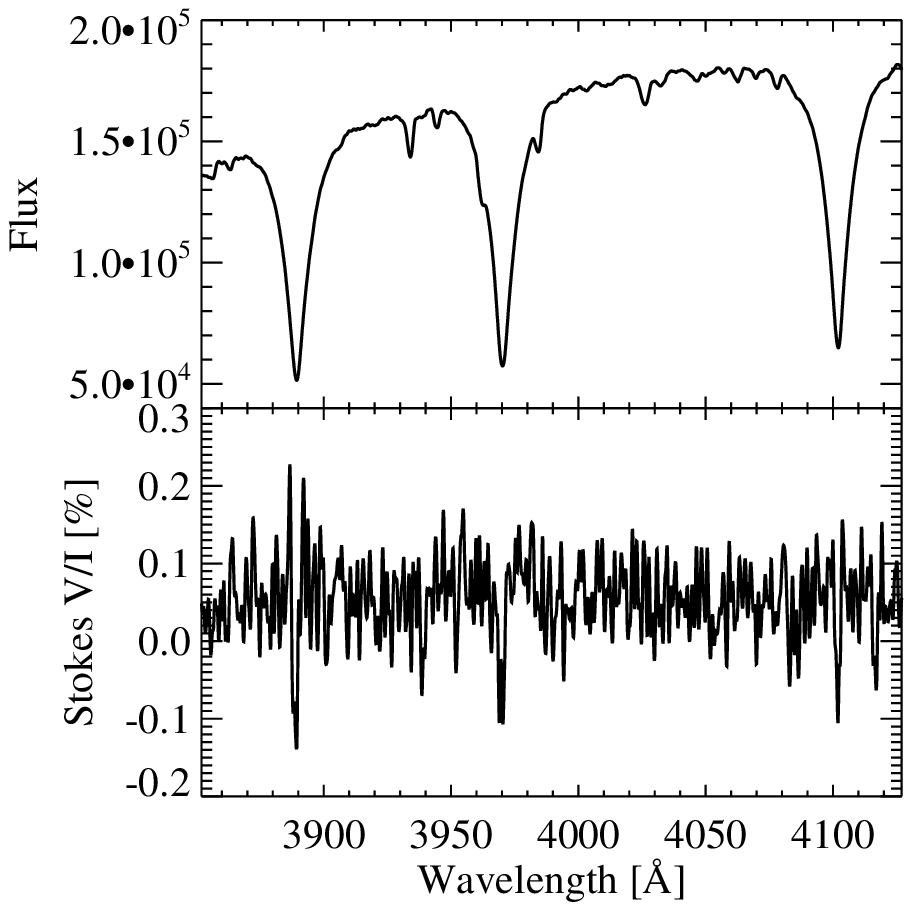}
\vskip-2mm
\caption{FORS1/2 observations of Stokes $I$ and Stokes $V$ spectra of the HgMn star HD\,65949. 
The \emph{left panel} corresponds to the data obtained in 2007, while 
the \emph{right panel} to the data from 2011. Distinct Zeeman features appear at the position of hydrogen lines
H8, H$\epsilon$, and H$\delta$. }
\label{FORS12}
\end{figure*}

A previous survey of magnetic fields in 17 HgMn stars (Hubrig et al.\ \cite{Hubrig2006b})
using low-resolution 
($R=2000$) circular polarization
spectra obtained with FORS\,1 at the VLT led to field detections in only four stars. 
This small sample of HgMn stars 
also included the spectrum variable HgMn star $\alpha$~And, for which a 
magnetic field of the order of a few hundred Gauss was detected. 
The HgMn star HD\,65949, for which a magnetic field was detected with FORS\,1 in 2006, was followed-up
by additional FORS\,1/2 observations in 2007 and in 2011. We were able to confirm the presence of the 
field with $\left<B_{\rm z}\right>=-116\pm32$\ G measured in 2007, and  $\left<B_{\rm z}\right>=-182\pm34$\ G
measured in 2011 (Hubrig et al., in preparation). It is of interest that the spectrum
of this star reveals the extremely large overabundance of Hg, Pt, Os, and especially Re (Cowley et al.\ \cite{Cowley2010}).  
In Fig.~\ref{FORS12} we present the most recent FORS\,1/2 Stokes $I$ and Stokes $V$ spectra of HD\,65949.

To further pinpoint the mechanism responsible 
for the surface structure formation in HgMn stars, we recently obtained
spectropolarimetric observations of AR\,Aur and investigated 
the presence of a magnetic field during a rotational phase of very 
good visibility of the spots with overabundant elements (Hubrig et al.\ \cite{Hubrig2010}). 
Since most elements are expected to be inhomogeneously distributed
over the surface of the primary of AR\,Aur,
magnetic field measurements 
were carried out for samples of Ti, Cr, Fe, and Y lines separately.
Among the elements showing line variability, the selected elements have numerous transitions 
in the observed optical spectral region, allowing us to sort out the best samples of clean
unblended spectral lines with different Land\'e factors.
A longitudinal magnetic field at a level higher than 3$\sigma$ of the order of a few hundred Gauss is detected
in Fe\,{\sc ii}, Ti\,{\sc ii}, and Y\,{\sc ii} lines, while a quadratic magnetic 
field ${\left<B\right>=8284\pm1501}$ G
at 5.5$\sigma$ level was measured in Ti\,{\sc ii} lines. No crossover at 3$\sigma$ confidence level was detected 
for the elements studied. Further, we detect a weak longitudinal magnetic field,
${\left<B_z\right>=-229\pm56}$ G, in the secondary component using a sample of nine Fe\,{\sc ii} lines.

The diagnosis of the 
quadratic field is more difficult than that of the longitudinal magnetic field, and it 
depends much more critically on the
number of lines that can be employed. Interestingly, the detection of a kG quadratic magnetic field in 
the primary and of a weak longitudinal magnetic field in 
the primary and the secondary of AR\,Aur is in accordance with results of Mathys \& Hubrig (\cite{Mathys1995})
who reported the detection of a 
weak magnetic field in the secondary of another SB2 binary, $\chi$\,Lup,
while in the same work they discovered a quadratic magnetic field of 3.6\,kG 
in the primary component of 74\,Aqr.
The secondaries in both, the AR\,Aur and $\chi$\,Lup, systems, are mentioned in the literature to have 
characteristics very similar to early Am stars.

\begin{figure*}
\centering
\includegraphics[trim=1.5cm 0cm 0.5cm 0cm,height=0.94\textwidth,angle=270]{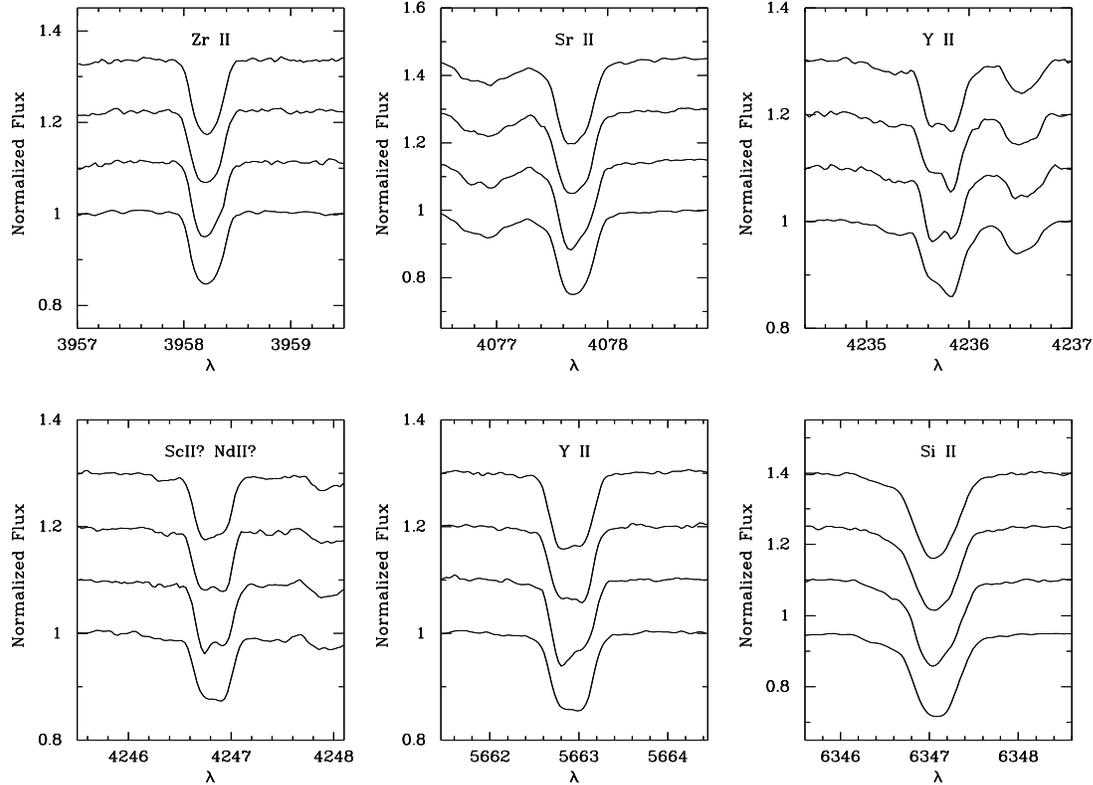}
\caption{Various degrees of variability  of line profiles in the 
spectra of HD\,101189 obtained with FEROS on four different nights.
For convenience, individual spectra are shifted in vertical direction.}
\label{fig_hd101}
\end{figure*}

The only longitudinal magnetic field measurements carried out for 
AR\,Aur were reported recently by Folsom et al.\ (\cite{Folsom2010}), who used the LSD (Least-Squares Deconvolution) technique
to combine 1168 lines of various elements. No magnetic field was detected in their analysis
of polarimetric spectra obtained in 2006. One possibility for this non-detection
could be related to an unfavorable element spot configuration or even to the absence of some element spots
at the epoch of their observations, since the authors report that no variability of the Ti and Fe lines was detected. 
It is of interest that strong Fe and Y element concentrations are almost missing in our first map based on SET1
observations in 2005, which are the closest in time to the  
observations reported by Folsom et al. New spectropolarimetric observations of AR\,Aur were recently obtained with 
the SOFIN spectropolarimeter installed at the 2.56-m Nordic Optical Telescope on La Palma. The current element distribution
appears somewhat different compared to our previous Doppler maps and the measured magnetic field is much weaker than 
that reported by us in 2010 (Hubrig et al., in preparation). Clearly, an extensive spectropolarimetric monitoring 
over several years is urgently needed to 
understand the puzzle of the dynamical evolution of chemical spots on the surface of these stars and the underlying 
structure of their magnetic fields.

Recently, Makaganiuk et al.\ (\cite{Makagan2011a}) reported their study of magnetic fields in HgMn stars using 
a polarimeter attached to the HARPS spectrograph at the ESO 3.6-m telescope. The authors used LSD profiles 
calculated for several hundreds of spectral lines. No detections at 3$\sigma$ level were reported 
for HgMn stars. Strangely enough, although the authors were aware of the 
possible inhomogeneous distribution  of elements on the surface of HgMn stars, no Zeeman signature
analysis was done on inhomogeneously distributed elements separately. Since a kind of symmetry
between the topology of magnetic fields and the element distribution is 
expected, the method of using all element spectral lines together is not advisable and
leads to doubtful results. Furthermore, 
their technique does not allow to measure other moments of the magnetic field. 

\phantom{ }

On the other hand, our own spectropolarimetric observations indicate that not every HgMn star possesses a 
magnetic field. As an example, the HgMn star HD\,71066 was observed by our team and by other groups, but no field 
was detected in any study (Hubrig et al.\ \cite{Hubrig1999b}, \cite{Hubrig2006b}; Makaganiuk et al.\ \cite{Makagan2011b}). 
Still, we do not know yet
whether the magnetic field is variable on secular time scales, as it appears to be the case for AR\,Aur.

To make ourselves familiar with HARPS polarimetric spectra, we downloaded from the ESO archive  in spring 2011 the 
Makaganiuk et al.\ publically available observations of the classical Ap star $\gamma$\,Equ and of the typical variable HgMn star HD\,101189.
The single-lined star HD\,101189 was already studied by us in the past with ESO instruments.
As an example of the variability of HD\,101189, we present in 
Fig.~\ref{fig_hd101} four spectra 
acquired on four different nights with FEROS on La Silla, where spectral 
lines of several elements exhibit various degrees of variability.

\begin{figure}
\vskip-3mm
\hskip-4mm
\includegraphics[width=85mm]{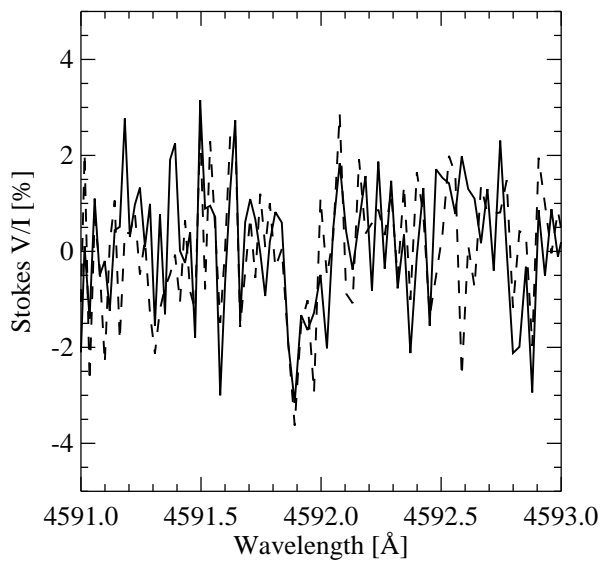}
\vskip-6mm
\caption{HARPS observations of HD\,101189. Stokes $V$ spectrum (solid line) and null spectrum (dashed line) are 
overplotted.}
\label{vz}
\end{figure}

\begin{table*}
\caption{HARPS observations.}
\label{HARPS}
\begin{tabular}{lcccc}\hline\noalign{\smallskip}
 Target & Date & Mod. Jul. Date & Signal-to-Noise & Numb.. of Spectra\\[1.5pt] 
\hline\noalign{\smallskip}
$\gamma$~Equ & 01-06-2009 & 54\,983.3703 & 97--216 & 4 \\
HD\,101189 & 31-05-2009 & 54\,982.9744 & 12--52\enspace & 4 \\[1.5pt]
\hline
\end{tabular}
\end{table*}

The details on the downloaded observations are listed in 
Table~\ref{HARPS}. Obviously, the quality of the spectra of HD\,101189 is especially poor with achieved 
signal-to-noise ratios (S/N) between 12 and 52. The data were reduced using the ESO HARPS pipeline. 
For the $\gamma$~Equ spectropolarimetric observations, which are usually used to
test the functionality of spectropolarimeters, the magnetic field measurements using the moment technique yield 
${\left<B_{\rm z}\right>=-904\pm68}$\,G and for the null spectrum ${\left<B_{\rm z}\right>=-3\pm15}$\,G.
The obtained longitudinal magnetic field fully agrees with the results of the FORS\,1 measurements by 
Hubrig et al.\ (\cite{Hubrig2004}). 

On the other hand, 
the measurements of the spectra of HD\,101189 deliver the same magnetic field (of the order of $-$200\,G) for both
the regular science Stokes $V$ observations and the null spectrum. As shown in 
Fig.~\ref{vz}, the reason 
for this is that the spectrum with the best S/N ratio (52 in this case) significantly contributes to the null spectrum. 
This result implies that if the observations are carried out with low S/N, and, 
in addition, if the difference in S/N in the sub-exposures 
is large, the magnetic field cannot be measured conclusively. The field can be measured rather accurately, 
but there is no use in the null spectrum to prove whether or not the detected field is real. 
Makaganiuk et al.\ (\cite{Makagan2011b}) 
report a non-detection with  ${\left<B_{\rm z}\right>=-95\pm66}$\ G calculating the LSD profile for the same spectrum.
As in the meantime additional HARPS spectra of HgMn stars became publically available, we plan their 
investigation in the near future.

\section{Conclusions}

Most stars in the studied sample of HgMn stars present a non-uniform distribution of several chemical elements. 
However, for only three stars the surface element distribution and its evolution over different time scales have been
studied until now. It is presently a fundamental question whether magnetic fields play a significant role in the 
development of abundance anomalies in HgMn stars, which are frequently members of binary and multiple systems. Answering this 
question is also important for the understanding of the processes taking place during the formation and evolution 
of B stars in multiple systems in general. A scenario describing how a magnetic field can be built up in binary systems was 
presented some time ago by Hubrig (\cite{Hubrig1998}) who suggested that a tidal torque varying with depth and latitude in 
a star induces differential rotation. Differential rotation in a radiative star can be prone to magneto-rotational 
instability (MRI). 

\begin{figure}
\centering
\includegraphics[width=0.50\textwidth]{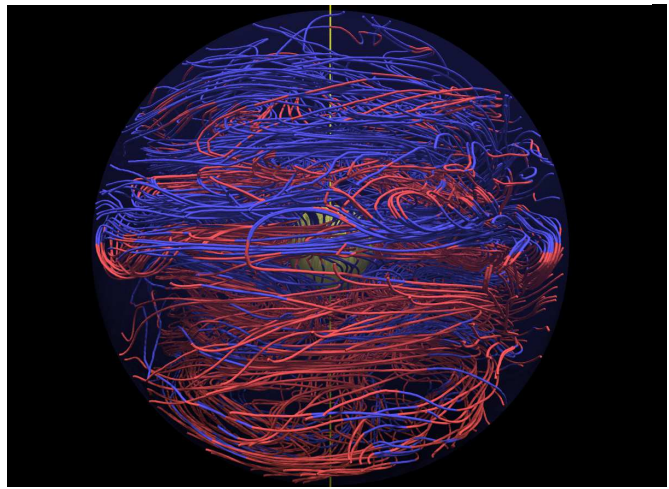}
\caption{(online colour at: www.an-journal.org) Magnetohydrodynamical simulations reveal a distinct magnetic field topology, which is similar to the 
fractured elemental rings observed on the surface of HgMn stars.}
\label{MRI}
\end{figure}

Magnetohydrodynamical simulations by Arlt, Hollerbach \& R{\"u}diger (\cite{Arlt2003}) revealed a distinct 
structure for the magnetic field topology similar to the fractured elemental rings observed on the surface of HgMn 
stars. The initial model differential rotation was hydrodynamically stable (Taylor-Proudman flow), but the introduction 
of a magnetic field excites the MRI on a very short time-scale compared to the time-scale of microscopic magnetic 
diffusion. Although fields are not very strong, complex surface patterns can be obtained from the nonlinear, 
nonaxisymmetric evolution of the MRI as shown in Fig.~\ref{MRI}. New simulations of the MRI in stellar radiation zones
are currently being performed and will deliver more details in the near future.
The aim of the future research projects is to search for a link between the magnetic 
nature and fundamental properties of a representative sample of 
spectroscopic binaries with late B-type primaries using spectroscopic and 
spectropolarimetric observations and the Magnetic Doppler Imaging technique.

\end{document}